\documentclass[lettersize,journal]{IEEEtran}
\usepackage{amsmath,amsfonts}
\usepackage{algorithmic}
\usepackage{algorithm}
\usepackage{array}
\usepackage{textcomp}
\usepackage{stfloats}
\usepackage{url}
\usepackage{verbatim}
\usepackage{graphicx}
\usepackage[inline,shortlabels]{enumitem}
\usepackage{cleveref}
\usepackage{siunitx}
\usepackage{subcaption}
\usepackage{multirow}
\usepackage{hhline}
\usepackage{nomencl}
\usepackage{enumitem}
\usepackage[style=ieee, maxnames=1, minnames=1]{biblatex}
\crefname{equation}{}{}
\Crefname{equation}{Eq.}{Eqs.}
\crefname{section}{Section}{Sections}
\Crefname{section}{Section}{Sections}
\crefname{figure}{Fig.}{Figs.}      
\Crefname{figure}{Figure}{Figures}  
\crefname{table}{Table}{Tables}
\Crefname{table}{Table}{Tables}

\makenomenclature
\begin{document}

\title{Model Predictive Control-Guided Reinforcement Learning \\ for Implicit Balancing}

\author{Seyed Soroush Karimi Madahi, Kenneth Bruninx,~\IEEEmembership{Member,~IEEE,}\\Bert Claessens, Chris Develder,~\IEEEmembership{Senior Member,~IEEE}
\thanks{Seyed Soroush Karimi Madahi and Chris Develder are with IDLab, Ghent University–imec, 9052 Ghent, Belgium (e-mail:\\ seyedsoroush.karimimadahi@ugent.be; chris.develder@UGent.be).

Kenneth Bruninx is with the Delft University of Technology, Faculty of Technology, Policy, and Management, 2600 Delft, GA, The Netherlands (email:k.bruninx@tudelft.nl).

Bert Claessens is with Beebop.ai, 2018 Antwerp, Belgium (email: bert@beebop.ai).}
}

\markboth{Journal of \LaTeX\ Class Files,~Vol.~14, No.~8, August~2021}%
{Shell \MakeLowercase{\textit{et al.}}: A Sample Article Using IEEEtran.cls for IEEE Journals}



\maketitle

\begin{abstract}
In Europe, profit-seeking balance responsible parties can deviate in real time from their day-ahead nominations to assist transmission system operators in maintaining the supply–demand balance. Model predictive control (MPC) strategies to exploit these implicit balancing strategies capture arbitrage opportunities, but fail to accurately capture the price-formation process in the European imbalance markets and face high computational costs. Model-free reinforcement learning (RL) methods are fast to execute, but require data-intensive training and usually rely on real-time and historical data for decision-making. This paper proposes an MPC-guided RL method that combines the complementary strengths of both MPC and RL. The proposed method can effectively incorporate forecasts into the decision-making process (as in MPC), while maintaining the fast inference capability of RL. The performance of the proposed method is evaluated on the implicit balancing battery control problem using Belgian balancing data from 2023. First, we analyze the performance of the standalone state-of-the-art RL and MPC methods from various angles, to highlight their individual strengths and limitations. Next, 
we show an arbitrage profit benefit of the proposed MPC-guided RL method of 16.15\% and 54.36\%, compared to standalone RL and MPC.
\end{abstract}


\nomenclature[A0]{{\textbf{Sets}}}{}
\nomenclature[A1]{$\hat{\Pi}_t$}{Set of indicative imbalance prices at time $t$}
\nomenclature[A2]{$\Omega$}{Set of uncertainty scenarios}
\nomenclature[A3]{$\text{aFRR}^{+/-}$}{Set of upward\,($+$)/downward\,($-$) aFRR bids}
\nomenclature[A3]{$\text{mFRR}^{+/-}$}{Set of upward\,($+$)/downward\,($-$) mFRR bids}
\nomenclature[A4]{$\text{LH}$}{Set of time steps in the look-ahead horizon}
\nomenclature[A5]{$\mathcal{D}$}{Experience replay buffer}

\nomenclature[B00]{{\textbf{Decision Variables}}}{}
\nomenclature[B01]{$\lambda_{t,\omega}$}{Imbalance price at time step $t$ for scenario $\omega$ [{\texteuro}/\,MWh]}
\nomenclature[B02]{$\lambda_{t, \omega}^\text{aFRR},\,\lambda_{t, \omega}^\text{mFRR}$}{aFRR, resp.\ mFRR, price at time $t$ for scenario $\omega$ [{\texteuro}/\,MWh]}
\nomenclature[B03]{$\textrm{SOC}_t$}{BESS state of charge at time $t$}
\nomenclature[B04]{$V_{t,\omega}^\text{aFRR},\,V_{t,\omega}^\text{mFRR}$}{Total activated aFRR, resp.\ mFRR, volume at time $t$ for scenario $\omega$ [MW]}
\nomenclature[B05]{$a_t$}{BESS action at time $t$ [MW]}
\nomenclature[B06]{$b_{t, \omega}^{a^{+/-}},\,b_{t, \omega}^{a^{+/-}}$}{Activated power of upward/downward regulation bid $a$ or $m$ at $t$ for scenario $\omega$ [MW]}
\nomenclature[B07]{$p_t^\text{dis},\,p_t^\text{cha}$}{BESS discharge, resp.\ charge, power at time $t$ [MW]}
\nomenclature[B08]{$\overline{p}^\textrm{bat}_{T_\textrm{qh}}$}{
Average BESS action within the quarter hour $\textrm{qh}$ before minute $T_\textrm{qh}$ [MW]}
\nomenclature[B09]{$r_t$}{RL reward at time $t$ [\texteuro]}
\nomenclature[B10]{$z_t^\text{BESS}$}{Binary variable to prevent simultaneous BESS charging \& discharging (1 = charging)}
\nomenclature[B11]{$z_{t, \omega}^\text{mFRR}$}{Binary variable indicating mFRR activation (1 = mFRR is activated)}

\nomenclature[C00]{{\textbf{Parameters}}}{}
\nomenclature[C01]{$\alpha$}{Entropy term weight in the actor loss}
\nomenclature[C02]{$\Delta t$}{Decision-making time resolution [h]}
\nomenclature[C03]{$\eta_\text{dis/cha}$}{BESS discharging/charging efficiency}
\nomenclature[C04]{$\mu$}{Soft update factor for the critic network}
\nomenclature[C05]{$\mu_t^{a^{+/-}},\,\mu_t^{m^{+/-}}$}{Activation price of upward/downward regulation bid $a$ or $m$ at time $t$ [{\texteuro}/\,MWh]}
\nomenclature[C06]{$\pi^\textrm{imb}_\textrm{qh}$}{Final imbalance price of quarter hour [{\texteuro}/\,MWh]}
\nomenclature[C07]{$B_t^{r^{+/-}}$}{Maximum power of upward/downward regulation bid $r$ for the quarter hour $t$ [MW]}
\nomenclature[C08]{$E_\text{b}$}{Maximum capacity of BESS [MWh]}
\nomenclature[C09]{$P_\text{BESS}$}{Maximum power of BESS [MW]}
\nomenclature[C10]{$\text{SI}_{t,\omega}$}{System imbalance at time $t$ for scenario $\omega$ [MW]}
\nomenclature[C11]{$\underline{\textrm{SOC}},\,\overline{\textrm{SOC}}$}{Minimum, resp.\ maximum BESS state of charge}
\nomenclature[C12]{$T_\textrm{qh}$}{Minute of the quarter hour $\text{qh}$ [min]}
\nomenclature[C13]{$\overline{V}_t^{\text{aFRR}^{+/-}}$}{Total upward/downward aFRR bid volume for the quarter hour $t$ [MW]}
\nomenclature[C13]{$\overline{V}_t^{\text{mFRR}^{+/-}}$}{Total upward/downward mFRR bid volume for the quarter hour $t$ [MW]}
\nomenclature[C14]{$m$}{Month of the year}
\nomenclature[C15]{$\textrm{qh}$}{Qaurter hour of the day}
\nomenclature[C16]{$s_t$}{RL state at time $t$}

\printnomenclature

\section{Introduction}
\IEEEPARstart{T}{he} intermittent nature of variable renewable energy sources (vRES) poses challenges to transmission system operators (TSOs) in maintaining the supply-demand balance. In Europe, the balancing responsibility has been partially outsourced to balance responsible parties (BRPs), such as energy companies~\cite{ENTSO}. In these settlement mechanisms, BRPs are responsible for maintaining the balance between consumption and generation within their portfolio. The integration of fast-response, flexible assets\,---\,such as battery energy storage systems (BESS)\,---\,along with improvements in forecasting techniques and decision-making tools, has enabled BRPs to deliberately take out-of-balance positions for financial profit, referred to as implicit balancing~\cite{smets2024participation}. Implicit balancing is a challenging problem due to nonlinearities, partial observability, and high uncertainty in the imbalance market. Various model-based optimization~\cite{smets2023strategic} and model-free reinforcement learning (RL) techniques~\cite{madahi2024distributional} have been explored in the literature to assist BRPs in implicit balancing actions.

Model-based optimization methods, such as model predictive control (MPC), are control strategies that use a market model and future forecasts to determine the optimal policy~\cite{garcia2021stochastic}. Despite the constraint enforcement and sample efficiency of these methods, they suffer from four major disadvantages:
\begin{enumerate*}[(i)]
    \item they require a precise electricity market model capturing the sub-quarter-hourly dynamics that govern price formation in European imbalance markets~\cite{madahi2025gaming}; 
    \item parametric and structural uncertainty complicate market modeling and decision making~\cite{dolanyi2023capturing}; 
    \item since they rely on solving an optimization problem on the fly, these methods can be computationally intensive, making them inefficient for real-time applications such as minute-level control of BESS~\cite{arroyo2022reinforced};
    \item their performance depends significantly on the accuracy of future predictions, such as system imbalance forecasts~\cite{hoang2023probabilistic}.
\end{enumerate*}

Model-free RL methods, as an alternative to MPC, do not require any forecaster or prior knowledge of the market environment~\cite{liang2020agent}. RL agents learn the optimal policy by interacting directly with the environment~\cite{boukas2021deep}. Although RL offers the advantages of being model-free and fast at inference, it faces two primary drawbacks:
\begin{enumerate*}[(i)]
    \item RL typically requires a large number of observations and extensive exploration to learn the (near-)optimal policy~\cite{zhu2024mpc};
    \item vanilla RL methods cannot guarantee the satisfaction of some constraints~\cite{pavirani2024demand}.
\end{enumerate*}

These advantages and disadvantages of RL and MPC highlight their complementary nature. To the best of our knowledge, there is no prior work that uses the combination of RL and MPC to control BESS in the imbalance settlement mechanism or the real-time market. Although combined approaches at different control levels have been widely studied in the literature (e.g.~\cite{fu2024novel}), few research works have explored the combination of MPC and RL for the same control problem~\cite{arroyo2022reinforced,gros2019data,li2021learning,bhardwaj2020blending}. Existing methods that merge RL and MPC for the same problem mostly either modify the MPC objective function by combining it with the RL value function~\cite{arroyo2022reinforced,bhardwaj2020blending}, or use MPC as a function approximator in RL~\cite{gros2019data,li2021learning}. However, these methods fail when the system model error is significant (as in the case of implicit balancing problems~\cite{smets2024participation}) since the executed control strategies are obtained by solving an optimization problem. These methods are thus also not appropriate for large-scale problems or for real-time applications due to their high computational time. 

We propose a novel MPC-guided RL method that can effectively use future forecasts for decision-making (as in MPC), while controlling the battery at a minute-level resolution (as in RL). We develop a stacked neural network architecture in which the output of the first neural network (RL-inspired), together with the MPC action and grid-related measurements, serves as input to the next network (the final decision-maker). The proposed architecture is trained end-to-end to maximize profit from implicit balancing using the RL framework. The performance of the proposed method is evaluated using Belgian imbalance prices from 2023. We benchmark the proposed MPC-guided RL method against standalone MPC and RL methods to highlight the benefits of their combination. 

Overall, our contribution is twofold:
\begin{enumerate}[leftmargin=0pt, labelindent=0pt, labelwidth=*, itemindent=0pt, align=left, nosep]
    \item For the first time, we compare the performance of state-of-the-art MPC and RL methods for implicit balancing from multiple perspectives\,---\,including profit and computational time\,---\,to identify their respective strengths and weaknesses. We investigate the impacts of look-ahead horizon, market model accuracy, and battery size on MPC performance by comparing results from deterministic MPC and RL. The effects of forecast quality on MPC performance, as well as the impact of different inputs and architectural choices on RL performance, are studied through comparing results from stochastic MPC and RL.
    \item We propose a novel MPC-guided RL method to leverage the complementary strengths of both MPC and RL. The proposed architecture consists of two stacked neural networks. The first neural network focuses on the RL state to extract useful information for decision-making. The second neural network uses the output of the first network and the MPC action, along with additional inputs\,---\,including forecast confidence inputs\,---\,to make the final decision. The proposed method effectively incorporates forecasts into the decision-making process through the MPC framework, which increases the expected profit, compared to standalone MPC and RL. Since the final decision maker is a neural network, the proposed method enables fast decision-making. Another advantage of the proposed method is its robustness to market model errors that deteriorate the performance of MPC actions. In cases of significant market model error, the proposed method bases its decisions on real-time balancing data, resulting in better performance.
\end{enumerate}

In the remainder of the paper, \cref{sec:related work}  first introduces the European imbalance settlement mechanism and outlines previous studies on implicit balancing as well as the combination of MPC and RL. \Cref{sec: methodology} explains our problem formulation and methodology in detail. Simulation results are presented in \cref{sec:results}, followed by the conclusion in \cref{sec:conclusion}.

\section{Background and Related Work}
\label{sec:related work}

\subsection{Imbalance Settlement Mechanism}
One of the TSO's main responsibilities is to maintain the demand-supply balance. To this end, frequency restoration reserve (FRR) volumes provided by Balancing Service Providers (BSPs) are activated in real time to correct system imbalances. At the end of each imbalance settlement period (ISP; typically 15\,min), BRPs are penalized or rewarded at imbalance prices that reflect the cost of these activated FRR volumes~\cite{lips2025insights}. This penalty or remuneration is computed based on the difference between BRP's day-ahead nominations and their real-time positions.

The European Electrical Balancing Guideline (EBGL) seeks to harmonize European imbalance settlement mechanisms, including aspects such as imbalance pricing methodology and ISP length (15\,min)~\cite{ENTSO}. All TSOs should apply single pricing methodologies, exposing BRPs with a short or long position to the same imbalance price. In this way, imbalance settlement remunerates BRPs that reduce the system imbalance while penalizing those that increase it. In this paper, we study the Belgian imbalance settlement mechanism, as it aligns closely with the target model in the EBGL: The Belgian TSO (Elia) adopts a single pricing methodology for each 15-minute ISP, in which imbalance prices are determined based on the sign of the total system imbalance and the volume-weighted average price of activated FRR~\cite{pavirani2025predicting}.

The EBGL allows BRPs to intentionally deviate from their schedule in order to earn imbalance remuneration. Some BRPs perform arbitrage across ISPs based on system imbalance forecasts for the next quarter hours. Some others adjust their position in response to real-time balancing data released by TSOs. This implicit balancing (also known as passive balancing or smart balancing) is not only beneficial for BRPs, but also promoted by some TSOs as it can reduce the system imbalance and the cost of FRR activation~\cite{doucet2023smart}: e.g., the Belgian TSO publishes near-real-time balancing data and the next quarter-hour system imbalance forecast to further reduce uncertainty and encourage BRPs to perform implicit balancing~\cite{Elia-forecast}. 

\subsection{Decision Support Tools for Implicit Balancing}
Some previous studies have focused on model-based optimization methods for implicit balancing or participation in the real-time market. Bottieau et al.~\cite{bottieau2019very} proposed a bi-level robust optimization problem to maximize the revenue of a BESS in the imbalance settlement. Li et al.~\cite{li2023ensemble} developed a robust short-term dispatch strategy for residential prosumers equipped with PV and a home battery, exposed to real-time prices, aiming to minimize household operating costs using ensemble nonlinear MPC. Lujano-Rojas et al.~\cite{lujano2016optimizing} introduced a comprehensive methodology to optimally control batteries operating in the real-time market, considering the effects of the charge controller operation, the variable efficiency of the power converter, and the maximum capacity of the electricity network. Smets et al.~\cite{smets2023strategic} proposed a stochastic MPC approach for implicit balancing for risk-averse BESS owners. 

The aforementioned optimization-based methods require an accurate and convex market clearing model. However, modeling the Belgian imbalance market model is complicated because
\begin{enumerate*}[(1)]
    \item the market model is non-convex, as the final price is calculated based on the volume-weighted average price of activated FRRs within each quarter hour; and
    \item uncertainty exists in FRR activations due to limited visibility on FRR provider availability and grid congestion, and is further compounded by the fact that imbalances are not always fully counteracted by the TSO (Elia)~\cite{allard2024forecast}.
\end{enumerate*}
Furthermore, the imbalance arbitrage problem formulation is non-linear. To limit the computational effort,  researchers often resort to linearization techniques (e.g., piecewise linear approximation), which may result in a poor approximation of the original problem. Since these methods require solving an optimization problem, they can be overly time-consuming during inference for real-time control tasks such as minute-based control of BESS. Finally, the performance of MPC is highly affected by the quality of the system imbalance forecaster used. Yet, predicting system imbalances is notoriously challenging due to their high uncertainty~\cite{van2024probabilistic}.

Reinforcement learning is a model-free alternative to MPC for implicit balancing in real-time markets~\cite{ruelens2016sequential}. In our previous work~\cite{madahi2024distributional}, we proposed an RL-based battery control framework for risk-sensitive implicit balancing, incorporating a cyclic constraint on the state of charge. In~\cite{karimi2024control}, we introduced an RL-based control framework for batteries to learn a safe energy arbitrage strategy based on defined human-intuitive properties in the imbalance settlement mechanism. Kwon et al.~\cite{kwon2022reinforcement} proposed an accurate cycle-based battery degradation model to develop efficient RL algorithms for the battery control problem in the real-time market. Although these data-driven methods achieve promising performance, they suffer from the sample efficiency issue, i.e., they need a large number of environment interactions to learn the (near-)optimal policy. Moreover, these models assume the BESS actions do not impact the final imbalance price.

Furthermore, although we demonstrated the importance of sub-quarter-hourly dynamics in setting the final imbalance price~\cite{madahi2025gaming}, none of the aforementioned research (based on MPC or RL) on implicit balancing takes into account the impact of these dynamics on their implicit balancing strategies. In this paper, we implement state-of-the-art MPC and RL methods to examine their respective advantages and limitations for implicit balancing strategies.

\subsection{Combining RL and MPC in other fields}
Despite the complementary nature of RL and MPC, their combination for solving the same control problem has rarely been studied in the literature. In~\cite{gros2019data}, economic nonlinear MPC schemes have been used instead of deep neural networks to support the parametrization of value functions and the policy in RL. Arroyo et al.~\cite{arroyo2022reinforced} proposed a reinforced MPC method that combines the MPC objective function with the RL agent value function while using a nonlinear controller model encoded from domain knowledge. Bhardwaj et al.~\cite{bhardwaj2020blending} presented a framework for combining MPC and RL to trade off model errors in the MPC and approximation error in a learned value function in RL by viewing MPC as constructing a series of local Q-function approximations. Zhu et al.~\cite{zhu2024mpc} introduced an MPC-guided RL scheme for an EV charging problem, where MPC serves as the baseline controller and RL is employed to learn control laws that compensate for model uncertainties and disturbances.

In all these methods, MPC serves as the primary controller and RL is used to mitigate system model errors or to extend the effective horizon of MPC. If the original MPC problem is time-consuming to solve, these methods cannot help reduce its computational time and may even increase it.

We propose an MPC-guided RL method in which the main decision maker is a stacked neural network trained using an RL framework. MPC actions are used as inputs to the proposed architecture to improve the sample efficiency. The proposed method is robust against system model errors because it learns the control strategy by interacting with the real imbalance market, with MPC actions provided solely as inputs to the proposed architecture. Furthermore, 
our method is fast at inference, as it uses a neural network for decision making. 

\section{Methodology}
\label{sec: methodology}

We seek an optimal control strategy for a profit-seeking BESS owner that performs implicit balancing. We formulate the decision problem both as 
\begin{enumerate*}[(i)]
\item a bi-level optimization problem to solve using MPC every quarter hour (\cref{subsec:optimization formulation}) and
\item a Markov decision process (MDP) to take action every minute, for which we will compare a vanilla RL and our newly proposed MPC-guided RL method (\cref{subsec:MDP formulation,subsec:mpc-guided rl}).
\end{enumerate*}

\subsection{Bilevel Optimization Problem}
\label{subsec:optimization formulation}
The upper level problem (\Cref{1,2,3,4,5,6,6b}) focuses on maximizing battery profit, and the lower level represents the balancing energy market clearing problem (\Cref{7,8,9,10,11,12,13,14,15}).
Thus, we capture the impact of battery actions on the imbalance price.

The objective function~\cref{1} maximizes battery profit in the imbalance settlement mechanism over the set of generated scenarios within defined look-ahead horizon. \Cref{2,3} define the update rule for the BESS state of charge (SoC) and specify its associated constraints. \Cref{4,5,6} ensure that the charging and discharging power in each quarter hour stays within the allowed maximum limits and prevent the BESS from charging and discharging simultaneously during that quarter hour. \Cref{6b} determines the imbalance price for the quarter hour obtained from the lower level problem.

\begin{equation}
   \max_{p_t^\text{cha},p_t^\text{dis}} \>\> \> \sum_{t \in \text{LH}}\sum_{\omega \in \Omega} \lambda_{t,\omega} \left( p_t^\text{dis} - p_t^\text{cha} \right) \Delta t
    \label{1}
\end{equation}
\vspace{-4ex}
\begin{align}
    &\text{Subject to:} \notag \\
    &\textrm{SOC}_{t+1} = \textrm{SOC}_t + \left(p_t^\text{cha} \> \eta_\text{cha} - \frac{p_t^\text{dis}}{\eta_\text{dis}} \right) \frac{\Delta t}{E_\text{b}} && \forall t \in \text{LH} \label{2} \\
    &\underline{\textrm{SOC}} \leq \textrm{SOC}_t \leq \overline{\textrm{SOC}} && \forall t \in \text{LH} \label{3} \\
    &0 \leq p_t^\text{cha} \leq z_t^\text{BESS} \> P_b && \forall t \in \text{LH} \label{4} \\
    &0 \leq p_t^\text{dis} \leq \left(1-z_t^\text{BESS}\right) \> P_b && \forall t \in \text{LH} \label{5} \\
    &z_t^\text{BESS} \in \{0,1\} && \forall t \in \text{LH} \label{6} \\ 
    &\lambda_{t,\omega} = \text{argmin}\left\{\text{\cref{7} s.t. \cref{8} -- \cref{15}}\right\} \hspace{2.6ex} \forall \omega \in \Omega, \hspace{-1.5ex} && \forall t \in \text{LH} \label{6b}
\end{align}

In the balancing energy market clearing process, TSOs minimize the cost of activating balancing energy for each quarter hour. 
\Cref{7} represents the objective of the imbalance market, which is to minimize the balancing activation cost. \Cref{8,9,10} impose constraints on aFRR and mFRR activations, with aFRR activation being prioritized over mFRR activation. The activation of necessary balancing energy is assured in~\Cref{11,12,13,14}. In line with previous studies \cite{smets2023strategic,bottieau2019very,smets2024participation}, we assume all scheduled frequency restoration control actions are executed perfectly. In other words, the frequency restoration control error is zero.
The imbalance price for the quarter-hour is determined as described in~\Cref{15}, following the Belgian implementation of the EBGL, as detailed below.

\begin{align}
   \min_{\substack{\omega \in \Omega, \\ t \in \text{LH}}}
   \quad & \sum_{a^+ \in \text{aFRR}^{+}} \mu_t^{a^+} b_{t,\omega}^{a^+}
   - \sum_{a^- \in \text{aFRR}^{-}} \mu_t^{a^-} b_{t,\omega}^{a^-} \notag \\
   &+ \sum_{m^+ \in \text{mFRR}^{+}} \mu_t^{m^+} b_{t,\omega}^{m^+}
   - \sum_{m^- \in \text{mFRR}^{-}} \mu_t^{m^-} b_{t,\omega}^{m^-}
   \label{7}
\end{align}
\vspace{-3ex}
\begin{align}
    &\text{Subject to:} \notag \\
    & \hspace{1ex} -\overline{V}_t^{\text{aFRR}^{-}} \leq V_{t, \omega}^\text{aFRR} \leq \overline{V}_t^{\text{aFRR}^{+}} && \hspace{-9ex}   \label{8} \\
    & \hspace{1ex} -\overline{V}_t^{\text{mFRR}^{-}} \!\! z_{t, \omega}^\text{mFRR} \leq V_{t, \omega}^\text{mFRR} \leq \overline{V}_t^{\text{mFRR}^{+}} \!\! z_{t, \omega}^\text{mFRR} && \hspace{-9ex}   \label{9} \\
    & z_{t, \omega}^\text{mFRR} = \begin{cases}
        1 &: V_{t, \omega}^\text{aFRR} = \overline{V}_t^{\text{aFRR}^{+}} \lor V_{t, \omega}^\text{aFRR} = -\overline{V}_t^{\text{aFRR}^{-}} \\
        0 &: \text{else}
    \end{cases} \label{10} \\
    & \hspace{3ex} V_{t, \omega}^\text{aFRR} + V_{t, \omega}^\text{mFRR} = p_t^\text{cha} - p_t^\text{dis} - \text{SI}_{t, \omega} && \hspace{-9ex}  
    \label{11} \\
    &\sum_{a^+ \in \text{aFRR}^{+}} \! \! \! \! b_{t, \omega}^{a^+}
   - \! \! \! \! \sum_{a^- \in \text{aFRR}^{-}} \! \! \! \! b_{t, \omega}^{a^-} \!=\! V_{t, \omega}^\text{aFRR} \hspace{0.5ex}: \lambda_{t, \omega}^\text{aFRR} && \hspace{-9ex}   \label{12} \\
    &\sum_{m^+ \in \text{mFRR}^{+}} \! \! \! \! b_{t, \omega}^{m^+}
   - \! \! \! \! \! \sum_{m^- \in \text{mFRR}^{-}} \! \! \! \! b_{t, \omega}^{m^-} \!=\! V_{t, \omega}^\text{mFRR} \hspace{0.5ex}: \lambda_{t, \omega}^\text{mFRR} && \hspace{-9ex}   \label{13} \\
   &\hspace{3ex} 0 \leq b_{t, \omega}^{r^{+}} \leq B_t^{r^{+}}, \quad
   0 \leq b_{t, \omega}^{r^{-}} \leq B_t^{r^{-}} && \hspace{-9ex}  \label{14} \\
    &\hspace{-2ex}\lambda_{t, \omega} = \begin{cases}
        \lambda_{t, \omega}^{\text{aFRR}} &: z_{t, \omega}^\text{mFRR}=0 \\
        \max(\lambda_{t, \omega}^{\text{aFRR}}, \lambda_{t, \omega}^\text{mFRR})  &: \text{SI}_{t, \omega} \leq 0 \land z_{t, \omega}^\text{mFRR}=1 \\
        \min(\lambda_{t, \omega}^{\text{aFRR}}, \lambda_{t, \omega}^\text{mFRR}) &: \text{SI}_{t, \omega} > 0 \land z_{t, \omega}^\text{mFRR}=1 \\
    \end{cases}
    \label{15}
\end{align}

The Belgian imbalance price for each quarter-hour is determined by automatic FRR (aFRR) and manual FRR (mFRR), which are activated according to the available regulation capacity merit order. During each quarter-hour, aFRR is activated through 4-second optimization cycles. The aFRR price for each direction in a given quarter-hour is calculated as the volume-weighted average price of the activated aFRR bids. In contrast, mFRR is manually activated by the operator either reactively (in response to the real-time system imbalance) or proactively (activated before the quarter-hour based on a forecast of future balancing needs). The most extreme activated mFRR bid (i.e., the highest price for upward regulation and the lowest for downward) sets the mFRR price for each direction in a given quarter hour. The final system imbalance of the quarter hour determines which FRR direction sets the final imbalance price. If both aFRR and mFRR are activated within a quarter hour, the more extreme price determines the imbalance price for that period~\cite{elia2023brp}. Other models available in the literature \cite{smets2023strategic,bottieau2019very,smets2024participation} do not consider these sub-quarter-hourly dynamics that determine imbalance energy prices. Furthermore, they consider only reactive mFRR activations.


MPC is a widely adopted method for solving control problems using a receding horizon approach. To deal with uncertainties, we use a scenario-based stochastic MPC approach~\cite{engels2019optimal}. At each time step $t$, the bi-level optimization problem defined above is solved to obtain the battery decision variables over the look-ahead horizon, considering a set of system imbalance forecasts. 

We transform the bi-level optimization problem into a mixed-integer linear program (MILP) by replacing the lower-level problem with its Karush-Kuhn-Tucker (KKT) conditions~\cite{saez2016data}. The non-anticipativity constraint is incorporated into the optimization problem by enforcing the battery decision variables to be scenario-independent over the whole look-ahead horizon. Finally, the obtained action for time step $t$ is applied to the battery. At time step $t+1$, after updating the battery's state of charge (SoC) and the inputs to the system imbalance forecaster, the process is repeated to determine the next battery action. 

Due to the limitations outlined in detail regarding imbalance market modeling, we adopt a quarter-hourly market model, in line with the state-of-the-art. Therefore, the time granularity for decision-making in MPC is 15\,min: at the beginning of each quarter hour, MPC determines the battery action for that entire interval.

\subsection{MDP Formulation}
\label{subsec:MDP formulation}
We can model the implicit balancing problem as a sequential decision-making problem using an MDP. The MDP problem is defined by a tuple $(\mathcal{S},\mathcal{A},\mathcal{R},\mathcal{P},\gamma)$, where $\mathcal{S}$ and $\mathcal{A}$ represent the state and action spaces, $\mathcal{R}: \mathcal{S} \times \mathcal{A} \rightarrow \mathbb{R}$ is the immediate reward function, $\mathcal{P}: \mathcal{S} \times \mathcal{A} \times \mathcal{S} \rightarrow [0,1]$ is the unknown state transition probability distribution, and $\gamma \in (0,1]$ denotes the discount factor~\cite{Sutton2018}.
At each time step, the state is defined as: 
\begin{equation}
    s_t=(T_\textrm{qh},\textrm{qh},m,\textrm{SOC}_t, \overline{p}^\textrm{bat}_{T_\textrm{qh}-1}, \hat{\Pi}_t) \quad,
    \label{rl-1}
\end{equation}
\begin{equation}
    \overline{p}^\textrm{bat}_{T_\textrm{qh}-1}= 
    \begin{cases}
       \sum_{t=0}^{T_\textrm{qh}-1} \frac{a_t}{T_\textrm{qh}}  &: T_\textrm{qh} \ge 1 \\
        0 &: T_\textrm{qh}=0 \\
    \end{cases} \quad,
    \label{rl-2}
\end{equation}
where $\overline{p}^\textrm{bat}_{T_\textrm{qh}-1}$ denotes the average BESS action within the quarter hour, calculated using~\Cref{rl-2}. $\hat{\Pi}_t$ is the set of indicative imbalance prices:
\begin{equation}
    \hat{\Pi}_t=\left\{\hat{\pi}_u|\hat{\pi}_u=f_\textrm{market}(b_{0:t-1}^{{a}^{+/-}}, b_{0:t-1}^{{m}^{+/-}},a_{0:t-1}, u),\, u\in\mathcal{A} \right\}
    \label{rl-3}
\end{equation}

Indicative imbalance prices are obtained using the activated aFRR/mFRR volumes and the battery actions taken during each quarter hour up to the minute $t$ of that quarter hour. They represent real-time imbalance prices based on cumulative balancing information within each quarter hour, up to minute $t$. Since the battery action influences FRR activations, these prices are calculated for all possible battery actions at minute $t$. These prices encapsulate useful information for the agent about the latest balancing conditions of the grid without violating the MDP property, as they are
computed solely based on historical data. However, these indicative prices are not the final imbalance prices.  The final imbalance price is only known at the end of the quarter hour and is calculated based on all activated FRR volumes within that quarter hour 
(see \cref{subsec:optimization formulation}).

The market model ($f_\textrm{market}(.)$) used to calculate indicative imbalance prices is more precise than the one described in~\cref{subsec:optimization formulation}: $f_\textrm{market}(.)$ is a nonlinear imbalance market model that considers the impact of minute-based FRR activation and reactive and proactive mFRR activations on calculated prices.\footnote{Elia determines aFRR prices using a 4-second optimization cycle. However, because balancing data (such as system imbalance and aFRR/mFRR activations) is published at one-minute intervals, this study adopts a one-minute resolution as the finest feasible granularity.} 

We define a discrete action space comprising three actions, expressed as follows:
\begin{equation}
    a_t \in \mathcal{A}, \, \mathcal{A}=\{-P_b, 0, P_b\} \, ,
    \label{rl-4}
\end{equation}
Battery charging is represented by a positive action, and discharging by a negative one. We adopt a discrete action space in this paper, motivated by the work of Seyde et al.~\cite{seyde2021bang}. However, it is worth mentioning that the proposed formulation and method in this paper can be naturally extended to a continuous action space.

As the agent aims to maximize arbitrage profit, the reward function is formulated as the negative of the imbalance cost: 
\begin{equation}
    r_t=-a_t \> \pi^\textrm{imb}_\textrm{qh}
    \label{rl-5}
\end{equation}

The state transition function $\mathcal{P}$, which models the system dynamics, is generally unknown, although the part related to the battery dynamics in our problem can be explicitly formulated using a linear battery model \cite{madahi2024distributional}. $\mathcal{P}$ depends on uncertainties in imbalance prices, system imbalances, and other market participants. The RL agent can implicitly estimate $\mathcal{P}$ through interaction with the environment.

The proposed RL agent seeks to learn a policy that maximizes the expected cumulative reward by solving the MDP problem introduced above. In our previous work~\cite{madahi2024distributional}, we concluded that distributional soft actor-critic (DistSAC) outperforms other RL methods in the context of energy arbitrage in the imbalance settlement mechanism. Hence, we control the battery every minute using DistSAC~\cite{ma2025dsac}. Distributional RL models the probability distribution over returns rather than a single expected return~\cite{bellemare2017distributional}. Incorporating a distributional perspective into RL offers several advantages, including enabling risk-sensitive policy learning, mitigating Q-value overestimation, and improving training stability.

The return distribution ($Z_\theta$) is estimated using a quantile distribution~\cite{dabney2018distributional}:
\begin{equation}
Z_\theta(s_t,a_t)=\sum_\text{i=1}^\text{N} q_{\tau_i}(s_t,a_t) \> \delta_{\tau_i}  \,, 
    \label{rl-6}
\end{equation}
where, $q_{\tau_i}$ represents the return corresponding to the $\tau_i$-th quantile, and $\delta_{\tau_i}$ denotes the Dirac delta function at $\tau_i$. The quantiles are uniformly spaced as $\tau_i = \frac{i}{N}$ for a fixed $N$.

In DistSAC, an actor network $\pi_\phi$ is trained to learn a policy that maximizes the Q-values estimated by a critic network $Z_\theta$, while also maximizing an entropy term to encourage exploration. The actor and critic networks are trained by minimizing their respective loss functions, $J_\pi$ and $L_Q$, defined as follows:
\begin{gather}
    J_\pi(\phi)=\mathbb{E}_{s \sim \mathcal{D}, a \sim \pi_\phi}\left[\alpha \ln \pi_\phi(a|s)-\mathbb{E}_{Z \sim Z_\text{$\theta$}}[Z(s,a)]\right] \\
    L_Q(\theta)=\mathbb{E}_{(s_t,a_t) \sim \mathcal{D}} \left[\sum_\text{i=1}^\text{N} \mathbb{E}_j[(\tau_i-1_{u_j<0})u_j]\right] \label{rl-7}\\
    u_j=\mathcal{T}q_{\tau_j}-q_{\tau_i}(s_t,a_t) \\
    \mathcal{T}q_{\tau_i}=r_t+\gamma \> \mathbb{E}_{a' \sim \pi_\phi} \left[q_{\tau_i}(s_{t+1},a')-\alpha \ln \pi_\phi(a'|s_{t+1})\right] \label{rl-8}\\
    \theta'=\mu \> \theta + (1-\mu) \> \theta' \qquad \mu \ll 1 \label{rl-9}  
\end{gather}

In~\Cref{rl-7}, the critic network is trained using the quantile regression loss. Quantile targets are calculated in~\Cref{rl-8} using the soft Bellman equation and a target critic network ($Z_{\theta'}$). For stability in training,  the parameters $\theta'$ are updated according to~\Cref{rl-9}, allowing the target network $Z_{\theta'}$ to slowly track $Z_\theta$~\cite{mnih2015human}.

\subsection{MPC-guided RL}
\label{subsec:mpc-guided rl}
\begin{figure}[t]
    \centering
    \includegraphics[width=\linewidth]{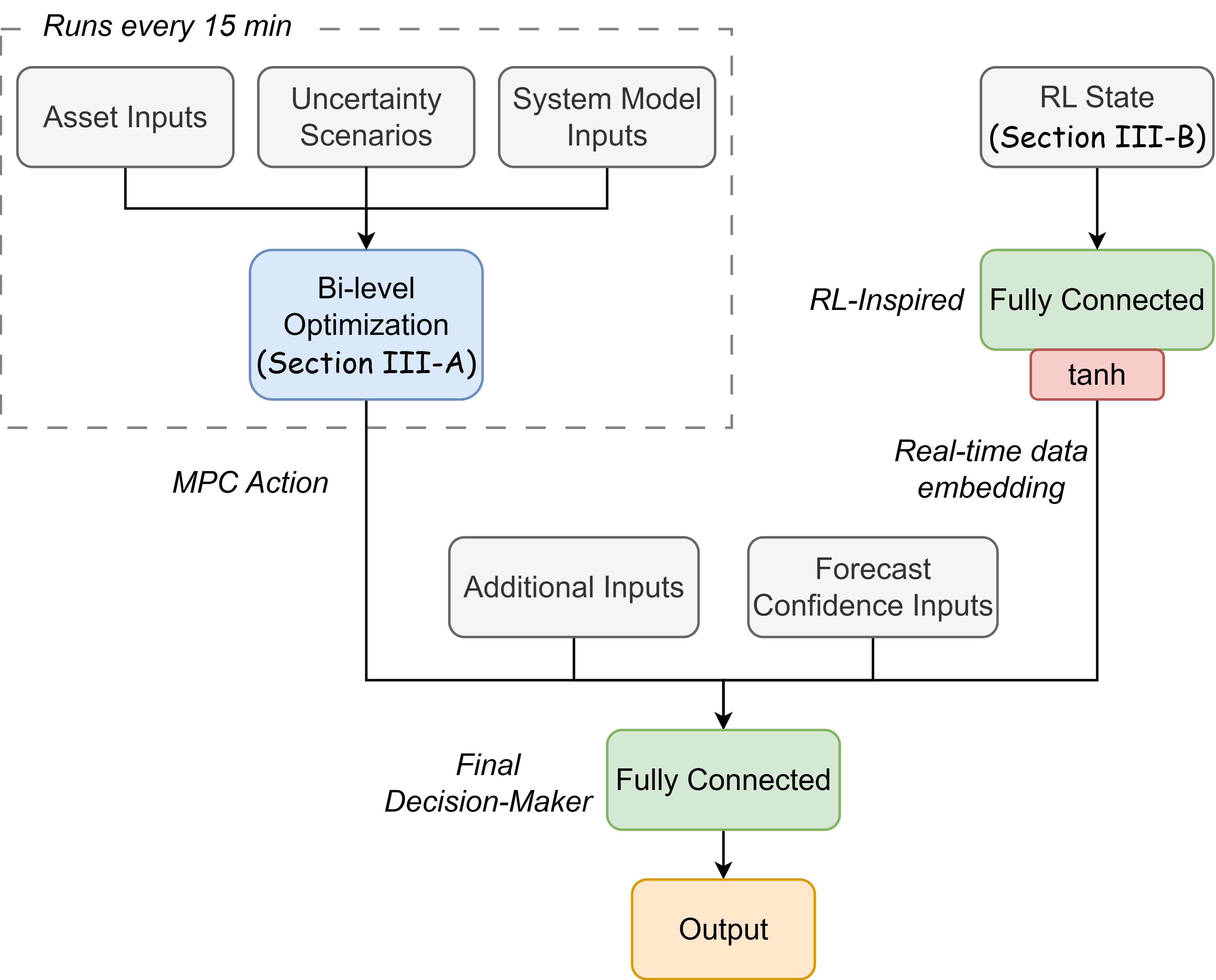}
    \caption{The proposed MPC-guided RL architecture. Real-time data (the RL state) is initially encoded by the RL-inspired network. The resulting embedding, the MPC action for that quarter hour and other inputs are then fed into the final decision-maker network to calculate the final action.}
    \label{fig:MPC-guided RL architecture}
\end{figure}

To combine the strengths of both MPC and RL, we propose a new MPC-guided RL architecture (\cref{fig:MPC-guided RL architecture}). The architecture consists of two stacked neural networks: an RL-inspired network and a final decision-maker network. The RL-inspired network encodes the RL state\,---\,including real-time data\,---\,into an embedding vector that captures essential features for decision-making. The proposed MPC-guided RL controls the BESS with a time resolution of one minute. At the beginning of each quarter hour, the MPC action for that quarter hour is calculated by solving the optimization problem defined in~\cref{subsec:optimization formulation}, and it remains fixed throughout the quarter hour.
The final decision-maker network's output is based on the resulting embedding vector along with the MPC action for that quarter hour, inputs related to forecast confidence, and other additional inputs.

Forecast confidence-related inputs help the final network better assess the quality of the MPC action. We use the minute of the quarter hour as the forecast confidence input, due to an inverse correlation between the minute of the quarter hour and the forecast confidence. For instance, at the last minute of a quarter hour, the total system imbalance up to that minute would be a better proxy for the final system imbalance over the quarter hour than the forecasted system imbalance at the beginning of the quarter hour. In other words, the trustworthiness of the forecasted system imbalance and the resulting MPC action would be low in the latter situation. Motivated by the work of Pavirani et al. \cite{Pavirani2025System}, we consider real-time activated aFRR and mFRR volumes as additional grid-related inputs to the final network to boost its performance.

Note that our proposed combination of MPC and RL differs from hierarchical methods in which MPC and RL collaborate at different control levels. In hierarchical methods, MPC typically determines a setpoint that RL tries to follow or each controller optimizes a different objective function. In our problem, both MPC and RL are used to solve the same problem at the same control level. More specifically, the MPC action is used as one of the inputs to the final decision-maker network, which has full freedom to deviate from the MPC action. The proposed MPC-guided RL architecture is trained end-to-end to solve the MDP problem defined in~\cref{subsec:MDP formulation}.

The proposed method improves upon standard RL by efficiently incorporating future forecasts. In~\cref{fig:MPC-guided RL architecture}, real-time data embedding captures current system information, while the MPC action encapsulates future forecasts. Forecast confidence inputs, along with other additional inputs, assist the final decision-maker network in determining which data to rely on more when taking the final action. In other words, these inputs modulate the influence of the encapsulated real-time and forecast data on the final action. On the other hand, the proposed method enhances MPC by being robust to system model errors. When the MPC action is suboptimal or inaccurate due to significant forecast errors or discrepancies in the system model, the proposed method can still rely on real-time data for decision-making and disregard the MPC action. The proposed architecture enables minute-level control of a BESS, 
allowing to exploit the imbalance price formation \cite{madahi2025gaming}.

\section{Results and Discussion}
\label{sec:results}
We study the performance of implicit balancing strategies using the Belgian balancing data of 2023. Each month's data is split into training (days 1–20), validation (days 21–25), and test (remaining days) sets. To study the impact of battery size, we consider four batteries with capacities of 1\,MW/\,2\,MWh, 10\,MW/\,20\,MWh, 50\,MW/\,100\,MWh, and 100\,MW/\,200\,MWh, each with a 90\% round-trip efficiency for both charging and discharging. The RL and MPC-guided RL agents are trained for \num{50000} episodes, with each episode simulating one day. The discount factor $\gamma$, the number of quantiles $N$, the soft update factor $\tau$, the experience replay buffer size, and the mini-batch size are set to \num{0.9995}, 20, 0.1, \num{1e6}, and \num{16384}. In vanilla RL, the actor and critic networks are modeled as fully connected neural networks, whereas in the proposed method, both are modeled using the architecture described in~\cref{fig:MPC-guided RL architecture}. The learning rates of the actor and critic networks are equal to \num{5e-5} and \num{5e-4}. To ensure the robustness of the results, all RL and MPC-guided RL agents are trained with five different seeds. The proposed architecture was implemented using PyTorch, and the optimization problem was solved using Gurobi. The experiments were conducted on a machine with a 6-core Intel Core i5 CPU (2.90\,GHz) and 32\,GB of RAM.

Our discussion is structured as follows. \cref{subsec:Exp deterministic MPC} illustrates the effect of battery size on the implicit balancing profit leveraging MPC and RL. The influence of the quality of system imbalance forecasts on MPC performance is discussed in \cref{subsec:Exp stochastic MPC}, whereas \cref{subsec:Exp MPC+RL} focuses on the impact of different inputs and architectural choices on RL performance. \cref{subsec:Exp MPC+RL} outlines the factors contributing to the superior performance of the proposed MPC-guided RL approach compared to standalone MPC and RL methods.


\subsection{Deterministic MPC vs.\ RL}
\label{subsec:Exp deterministic MPC}

\begin{figure}[t]
    \centering
    \begin{subfigure}{0.41\textwidth}
        \includegraphics[width=0.95\linewidth]{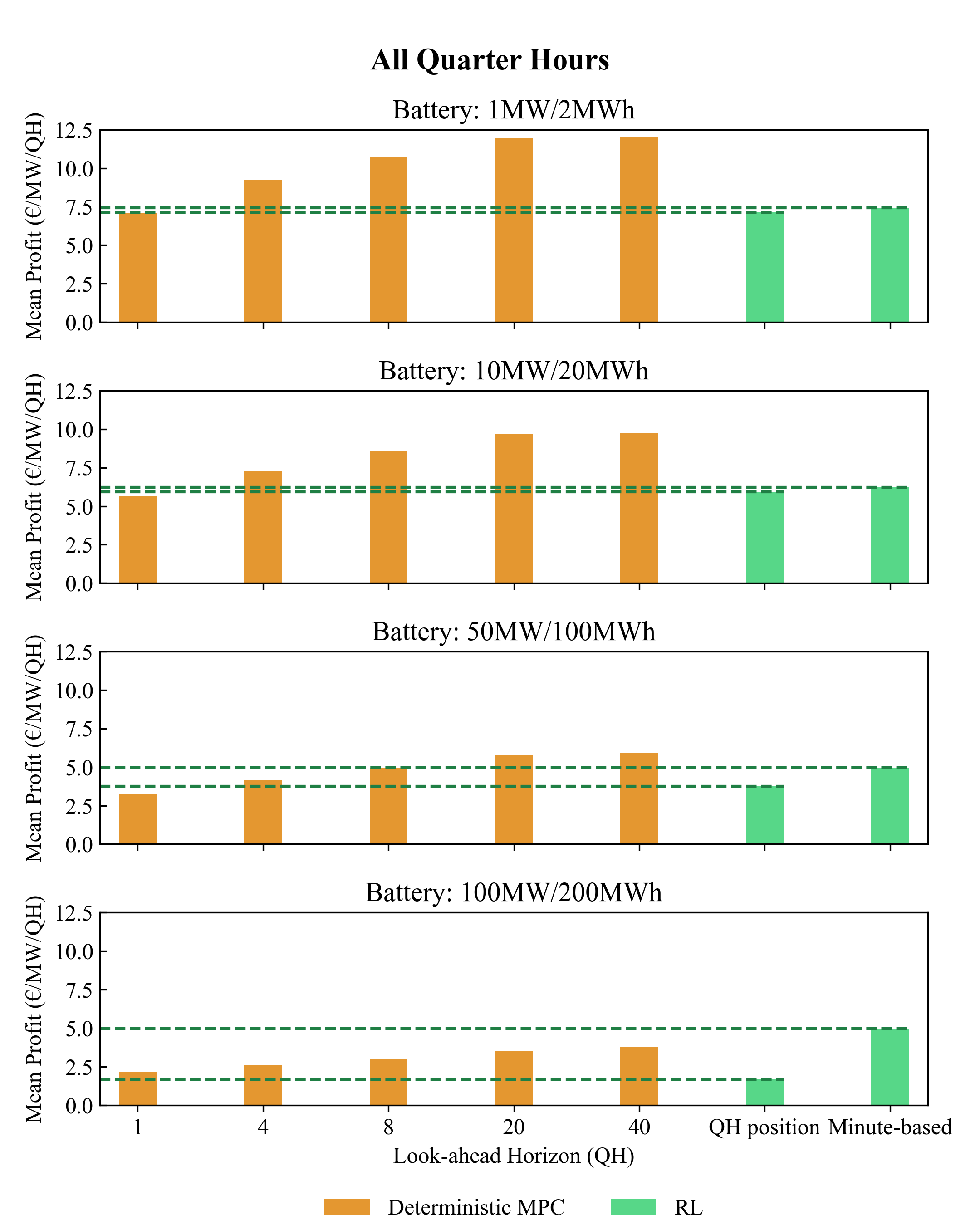}
        \caption{}
        \label{fig:RL_vs_Deterministic_MPC_total}
    \end{subfigure}
    \begin{subfigure}{0.41\textwidth}
        \includegraphics[width=0.95\linewidth]{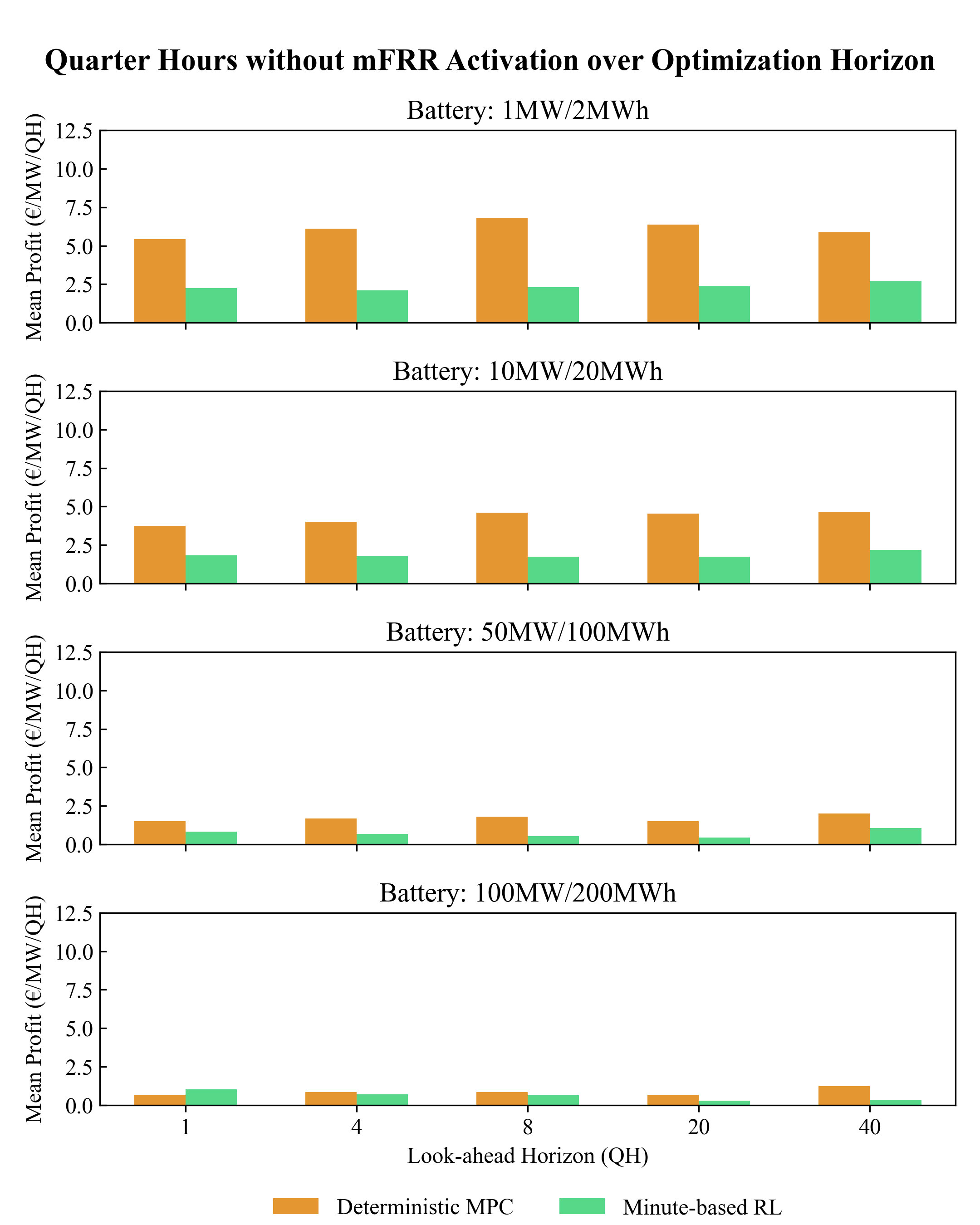}
        \caption{}
        \label{fig:RL_vs_Deterministic_MPC_no_mfrr_horizon}
    \end{subfigure}
    \caption{The deterministic MPC vs.\ RL results for (a)~all quarter hours and (b)~quarter hours without mFRR activation over the optimization horizon. In (b), RL results vary across different look-ahead horizons because of differently selected quarters for each horizon.
    }
    \label{fig:RL_vs_Deterministic_MPC}
\end{figure}

To study the impact of battery size on arbitrage revenue, we compare the performance of the deterministic MPC with perfect foresight across different look-ahead horizons to that of the base RL agent formulated in~\cref{subsec:MDP formulation}. \Cref{fig:RL_vs_Deterministic_MPC} shows that, as expected, increasing battery power leads to a reduction in BRP profit due to the greater impact of large batteries on the imbalance market. By lowering the volume of activated FRR, large batteries further reduce imbalance prices, resulting in lower BRP profits. Longer look-ahead horizons allow arbitraging across different imbalance settlement periods and strategically selecting quarter hours to offer implicit balancing, resulting in increased profits.

As battery power increases, the profit of the deterministic MPC decreases more sharply compared to that of the minute-based RL. For the 100\,MW battery, the minute-based RL agent (which makes decisions solely based on real-time data in~\Cref{rl-1}) outperforms the deterministic MPC with a 40-quarter-hour look-ahead horizon. Two effects are at play.

First, the minute-based RL agent takes battery actions every minute, which allow exploiting the imbalance pricing structure \cite{madahi2025gaming}, outperforming the MPC for larger batteries. To highlight the importance of the sub-quarter-hourly dynamics in implicit balancing, we compute the equivalent QH-position based on the average action of the minute-based agent over that quarter hour and apply this as a constant action over the entire quarter hour. The deterministic MPC outperforms this benchmark in all cases, provided the look-ahead horizon 
spans at least 4 quarter hours. The superior performance of the minute-based RL agent over the QH-position agent underscores the potential benefits of adjusting MPC battery actions within the quarter hour, especially for large batteries. 

Second, market model inaccuracies in the MPC formulation lead to poor control action decisions. Recall that we omitted proactive mFRR activations in the convex market model, whereas the RL agent is able to infer these from interactions with the environment. As shown in~\cref{fig:RL_vs_Deterministic_MPC_no_mfrr_horizon}, during quarter hours \textit{without} proactive mFRR activation in the optimization horizon, the deterministic MPC always achieves higher profit than the minute-based RL, as expected. This underscores the potential benefits of adjusting MPC battery actions via RL to correct for market model errors, especially for large batteries.   

\subsection{Stochastic MPC vs.\ RL}
\label{subsec:Exp stochastic MPC}

We examine how system imbalance forecast error affects the performance of MPC relative to RL. For this purpose, we add zero-mean Gaussian noise to the real system imbalance values, allowing us to control the forecast error by adjusting the noise's standard deviation. We consider an exponential growth of 20\% for $\sigma$ across the forecast horizon, based on the intuition that forecasts further into the future are less accurate. As a benchmark for currently achievable forecast performance, we adopt the state-of-the-art probabilistic system imbalance forecaster proposed in~\cite{van2024probabilistic} to evaluate the performance of the stochastic MPC equipped with a realistic forecaster. The forecaster outputs quantile values for quarter hour intervals within the look-ahead horizon. We use these quantiles to sample system imbalances for the corresponding quarter hours. The stochastic MPC results are obtained using 20 sampled system imbalance scenarios per quarter hour, with a look-ahead horizon of 4 quarter hours.

\begin{figure}[t]
    \centering
    \begin{subfigure}{0.45\textwidth}
        \includegraphics[width=\linewidth]{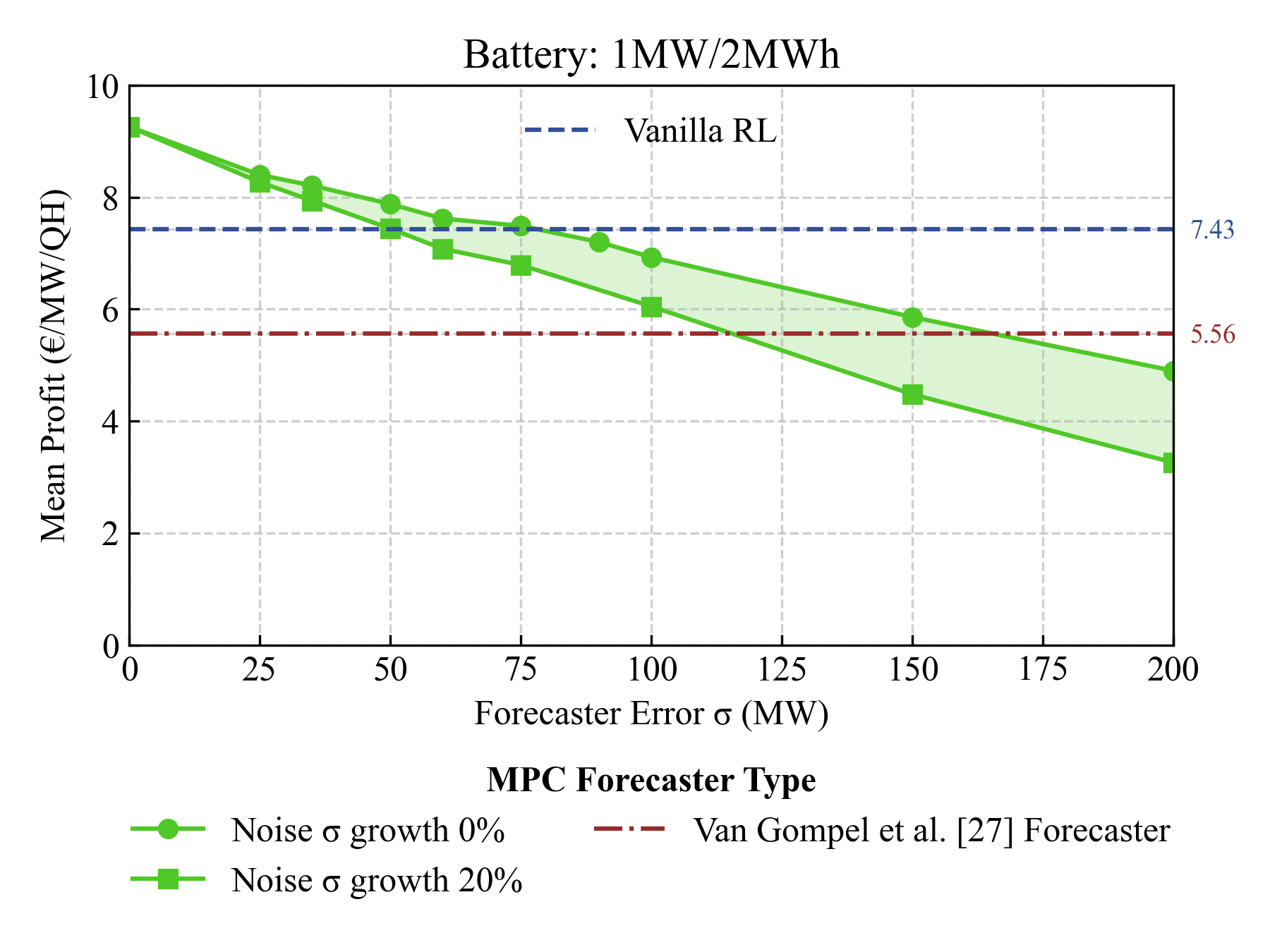}
        \caption{}
        \label{fig:RL_vs_stochastic_MPC_1MW}
    \end{subfigure}
    \begin{subfigure}{0.45\textwidth}
        \includegraphics[width=\linewidth]{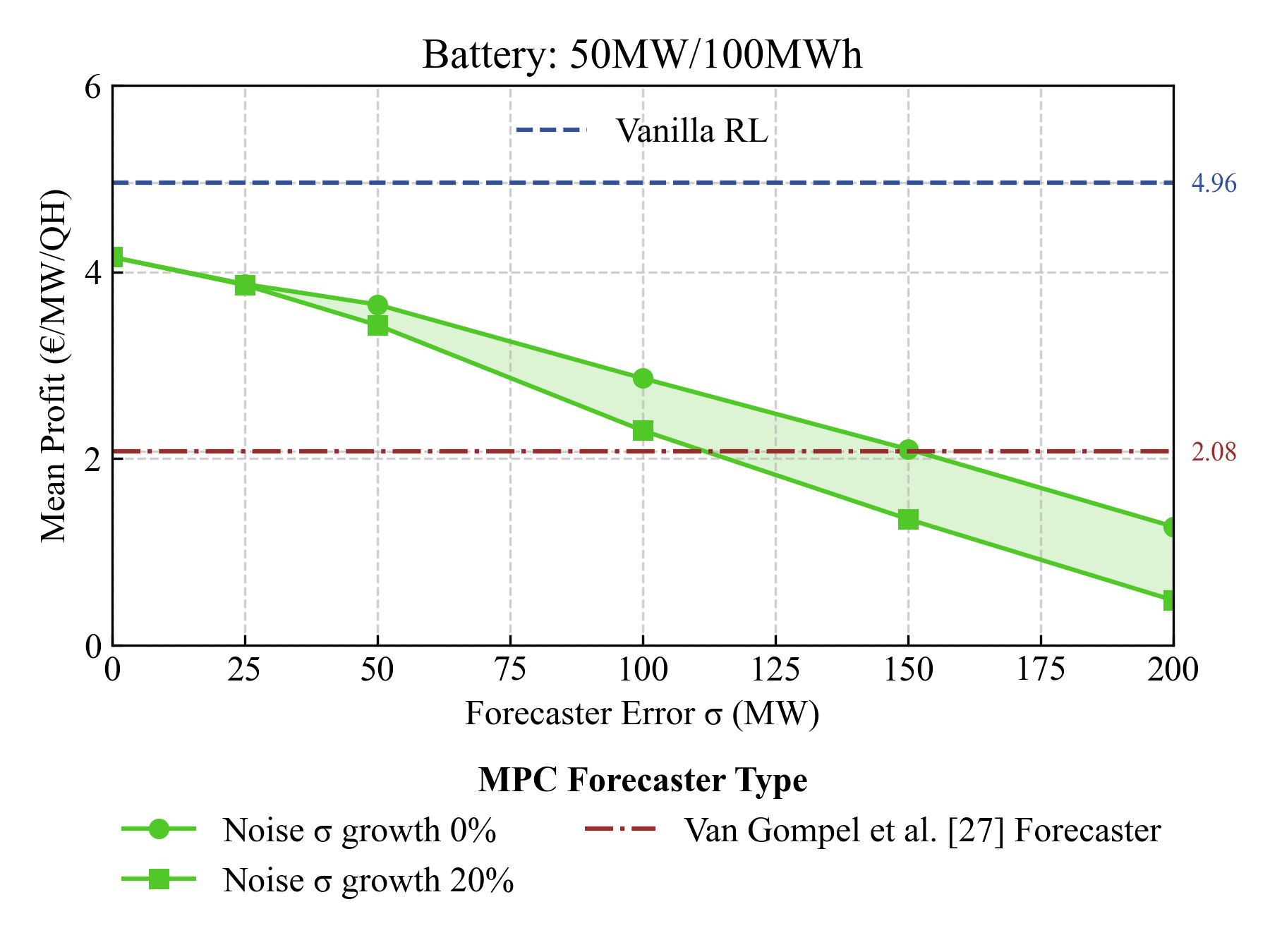}
        \caption{}
        \label{fig:RL_vs_stochastic_MPC_50MW}
    \end{subfigure}
    \caption{The stochastic MPC vs.\ RL results for (a)~the 1\,MW battery and (b)~the 50\,MW battery.}
    \label{fig:RL_vs_stochastic_MPC}
\end{figure}

\Cref{fig:RL_vs_stochastic_MPC_1MW} shows that stochastic MPC and RL achieve comparable performance when the system model used in MPC is accurate. The MPC performance is closely tied to the quality of the forecasts (i.e., the generated scenarios). Vanilla RL surpasses stochastic MPC with the state-of-the-art system imbalance forecaster by 33.6\% (1MW BESS) and 138.5\% (50MW BESS). The larger gap between RL and MPC for the 50\,MW battery in~\cref{fig:RL_vs_stochastic_MPC_50MW} is due to the compounding of market model errors with system imbalance forecast uncertainty.

\Cref{table:runtime} presents the computational time of stochastic MPC with a 4-quarter-hour look-ahead horizon using the real forecaster and vanilla RL for the two batteries over 10 sample days. The results reveal that increasing the number of scenarios from 20 to 100 for the small battery improves arbitrage profit by 19.7\%, while raising the runtime by $\sim$20$\times$. 
Aside from the mathematical complexity of incorporating sub-quarter-hourly dynamics into the imbalance market model, stochastic MPC is not well-suited for minute-based battery control due to its high computational cost, especially for large batteries. For the 50\,MW battery, solving stochastic MPC using 20 scenarios takes over half a minute, making it unsuitable for minute-based battery control\,---\,especially compared to RL, which requires only 0.024\,s to compute the next action.

\begin{table}[h!]
\centering
\caption{Mean profit and runtime of stochastic MPC and RL for various battery sizes over the last 10 days of January 2023.}
\begin{tabular}{ccccc}
\hhline{=====}
\textbf{Battery} & \textbf{Method} & \textbf{Scenario \#} & \begin{tabular}{@{}c@{}} \textbf{Profit} \\ \textbf{(\texteuro/\,MW/\,QH)}\end{tabular}  & \begin{tabular}{@{}c@{}} \textbf{Runtime} \\ \textbf{(s/\,QH)}\end{tabular} \\ \hhline{=====}
\multirow{5}{*}{1\,MW/\,2\,MWh} & \multirow{4}{*}{\begin{tabular}{@{}c@{}} Stochastic \\ MPC \end{tabular}} & 20 & 3.2 & 3.13\\
 & & 30 & 3.23 & 5.73\\
 & & 50 & 3.42 & 17.08\\
 & & 100 & 3.83 & 68.51\\
 & \textbf{RL} & -- & \textbf{5.98} & \textbf{0.024}\\ \hline
\multirow{5}{*}{50\,MW/\,100\,MWh} & \multirow{4}{*}{\begin{tabular}{@{}c@{}} Stochastic \\ MPC \end{tabular}} & 20 & 0.94 & 37.16\\ 
 & & 30 & 0.95 & 63.78\\
 & & 50 & 1.07 & 117.09\\
 & & 100 & 1.09 & 458.68\\
 & \textbf{RL} & -- & \textbf{3.28} & \textbf{0.024}\\ \hhline{=====}
\end{tabular}
\label{table:runtime}
\end{table}

\subsection{Proposed MPC-Guided RL}
\label{subsec:Exp MPC+RL}

We compare the performance of the proposed MPC-guided RL agent against that of three RL agents with a fully connected neural network architecture: 
\begin{enumerate*}[(1)]
    \item ``vanilla RL'': agents without any forecast-related input, but considering current minute-based FRR activation volumes (``with FRR'') or exclude this information (``base'');
    \item ``RL+forecast'': agents with system imbalance forecasts as inputs;
    \item ``RL+MPC FC'': agents with the quarter-hour MPC action with a look-ahead horizon of 4 quarter hours as an input.
\end{enumerate*}
We consider a 1\,MW and a 50\,MW battery. For the 1\,MW battery, the performance of stochastic MPC and RL is comparable (Section \ref{subsec:Exp stochastic MPC}) and the convex imbalance market model accurately reflects imbalance price formation as the battery is not large enough to influence FRR activations. As the latter does not hold for the 50\,MW battery, we use this experiment to test the robustness of the proposed method w.r.t. suboptimal MPC actions.

The proposed MPC-guided RL method outperforms all benchmarks for the 1\,MW battery, regardless of the forecaster type used in the MPC (\cref{fig:rl comparison_1MW}). The proposed method (blue box plots) with the real forecaster achieves profit improvements of 16.15\% and 54.35\% over the base RL agent (red box plots) and stochastic MPC (horizontal lines). These results confirm the complementarity between RL and MPC. While more accurate forecasts lead to better performance overall due to better MPC actions (blue boxplot, $\mathcal{N}(0,50)$), even MPC agents with lower performance than the RL agent can still enhance the performance of an RL agent (other blue boxplots). On the other hand, the 50\,MW battery results (\cref{fig:rl comparison_50MW}) reveal that adding MPC actions leads to only a 5.2\% improvement over the base RL agent, because of the low quality of MPC actions and significant market model errors. The differences in performance can be attributed to three factors:

\begin{figure}[t]
    \centering
    \begin{subfigure}{0.49\textwidth}
        \includegraphics[width=\linewidth]{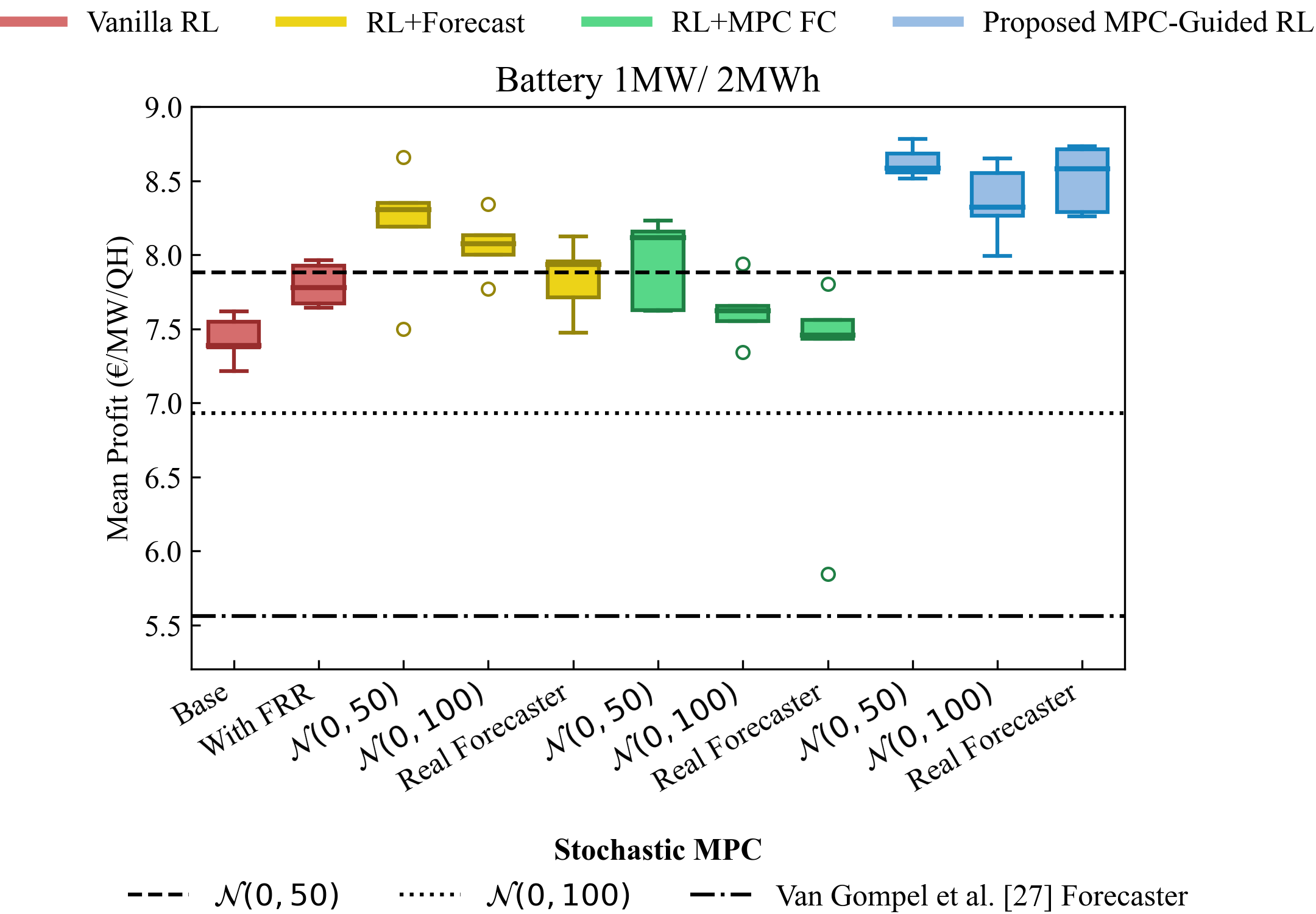}
        \caption{}        
        \label{fig:rl comparison_1MW}
    \end{subfigure}
    \begin{subfigure}{0.49\textwidth}
        \includegraphics[width=\linewidth]{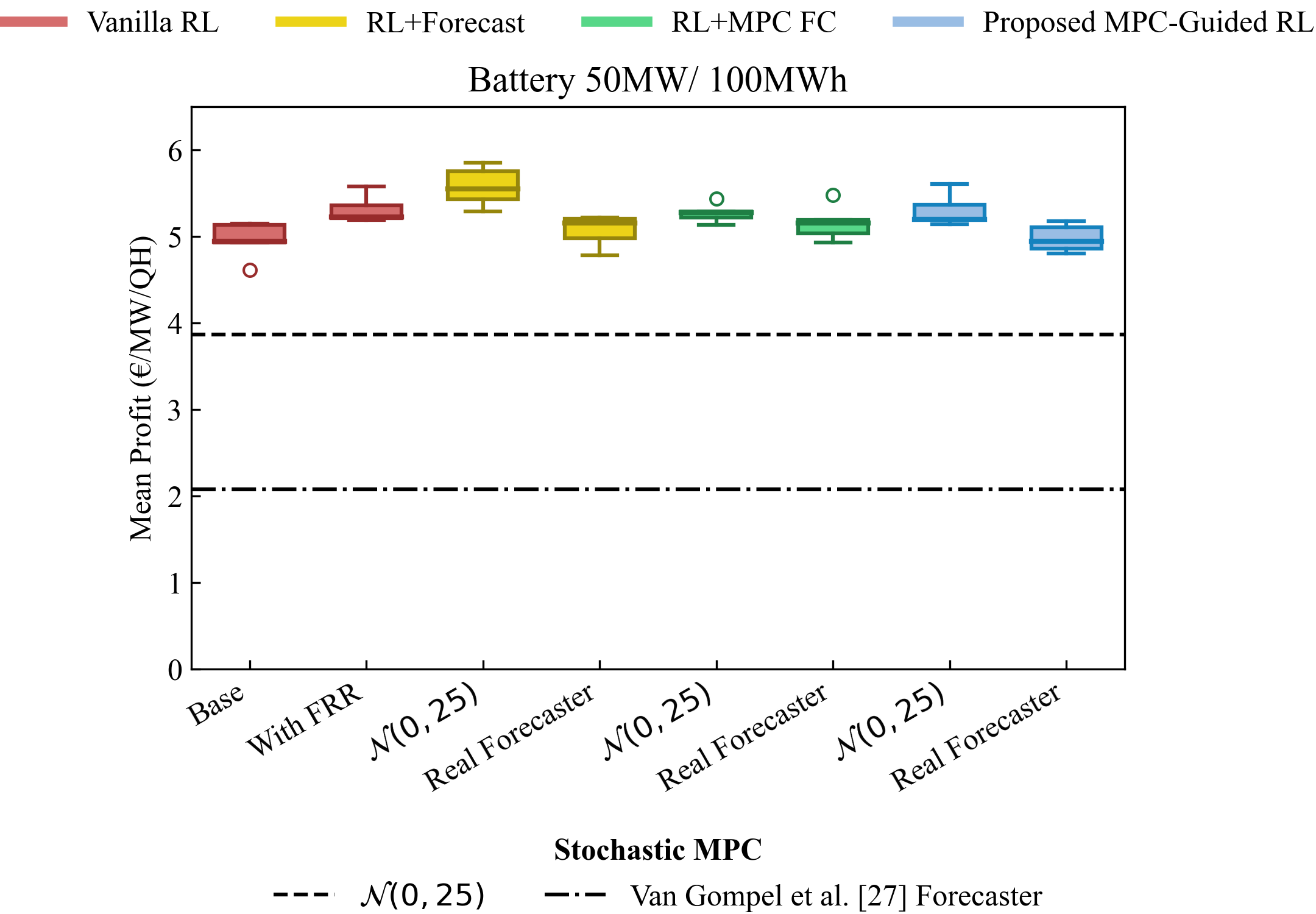}
        \caption{}        
        \label{fig:rl comparison_50MW}
    \end{subfigure}
    \caption{The comparison of various trained RL agents for (a)~1\,MW and (b)~50\,MW batteries.}
    \label{fig:rl comparison}
\end{figure}

\paragraph{Current FRR activation as input} The RL agent which includes FRR activation inputs (right red boxplot) outperforms the vanilla RL (left red boxplot) agent by 5.3\%. As FRR activation volumes directly influence the imbalance price, this offers the RL agent a better expectation of system conditions and imbalance prices.
\paragraph{System imbalance forecasts and MPC actions as input} The proposed MPC-guided RL agents (blue boxplots) and the RL+forecast agents (yellow boxplots) achieve higher performance than the base RL agent (red boxplot). These system imbalance forecasts inform the RL agent about to-be-expected system conditions and imbalance prices. By leveraging a MPC strategy to translate these forecasts in imbalance prices across a longer look-ahead horizon, MPC-guided RL agents (blue boxplots) exceed mean profit of the RL+forecast agents (yellow boxplots) by 4.87\% for the 1\,MW battery. This indicates that the optimization block in~\cref{fig:MPC-guided RL architecture} can powerfully encapsulate useful information from the input system imbalance forecasts into a single value (the MPC action).  

\begin{figure}[t]
    \centering
    \includegraphics[width=0.7\linewidth]{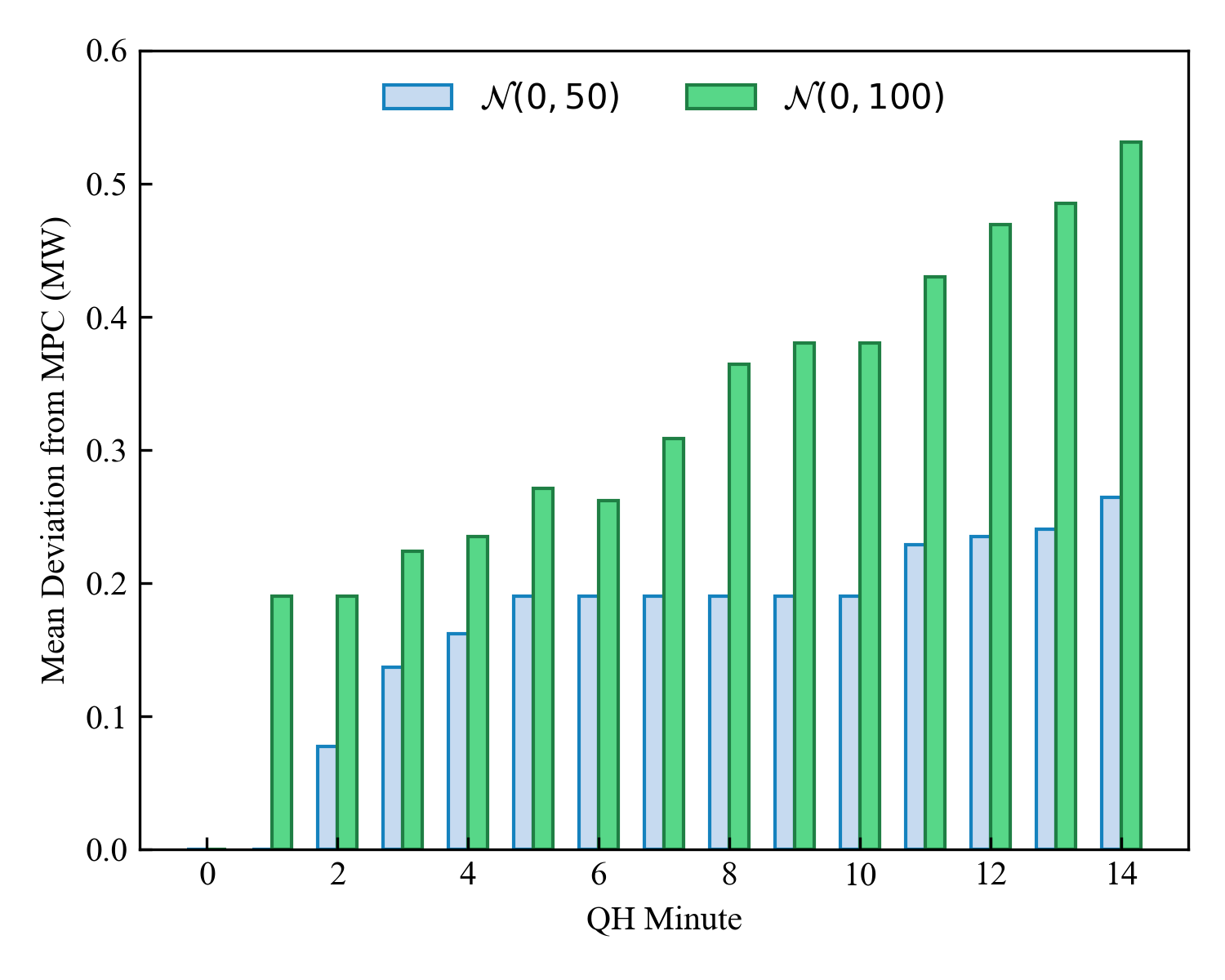}
    \caption{The average deviation of the proposed MPC-guided RL agent actions from the MPC actions for the 1\,MW battery under different forecast errors.}
    \label{fig:RL_action_deviation}
\end{figure}

\Cref{fig:RL_action_deviation} highlights how the proposed MPC-guided RL agent leverages forecasts (embedded in the MPC) and real-time balancing data in determining the battery dispatch. At the beginning of each quarter hour, the agent follows the MPC action, hence, relies more on predictions than on real-time information. As the quarter hour unfolds, the agent deviates from the MPC action, since real-time balancing data becomes a better prediction of the system imbalance and imbalance prices. Also, the proposed agent relies more on MPC actions as forecast accuracy increases. The proposed agent implicitly learned the quality of the forecasters used in MPC, resulting in greater deviations from the MPC actions when less accurate forecasts are used.

However, in the presence of significant market model error, MPC actions can mislead the RL agents, as illustrated for the 50\,MW battery in \cref{fig:rl comparison_50MW}. In such cases, the RL agents (yellow boxplots) are able to extract useful information from the raw forecast data more effectively than the optimization model (blue boxplots).
\paragraph{The NN architecture} The impact of the architecture design is evident from the performance of the RL+MPC FC agents in which the proposed architecture (blue boxplots) is replaced with a fully connected one (green boxplots). The performance of the latter is more sensitive to the forecast quality in the MPC. The proposed MPC-guided RL agents achieve better results than the RL+MPC FC agents, especially for the 1\,MW battery. In contrast to the FC architecture, the proposed architecture separates (i) grouping all related inputs and feature extraction and (ii) decision-making. 


\section{Conclusion}
\label{sec:conclusion}
We proposed an MPC-guided RL method that effectively leverages the advantages of both Model Predictive Control (MPC) and Reinforcement Learning (RL) to guide implicit balancing actions of BESS owners. The proposed method consists of two stacked neural networks: the first one encodes real-time data (the RL state), and the second one outputs the final action using the embedded real-time data, the MPC action for that quarter hour, and other additional inputs. 

The proposed MPC-guided RL control strategy leverages the complementary strengths of RL and MPC. MPC allows leveraging forecasts to optimize battery actions over a (long) look-ahead horizon, but is computationally intractable for minute-level control, relies on convex approximations of the imbalance market, and requires strong assumptions on the behavior of the system operator. RL is fast at inference, but requires large datasets for training and may lead to lower profits, especially for smaller BESS and/or when accurate system imbalance forecasts are available.  


The proposed MPC-guided RL method outperforms standalone RL and stochastic MPC with the state-of-the-art system imbalance forecaster by 16.15\% and 54.36\% in terms of profit in an implicit balancing problem for a 1\,MW battery using Belgian balancing data from 2023. Similar performance gains were not observed for larger batteries, as their impact on the imbalance price formation is not well captured in the MPC, resulting in low-quality MPC actions. The results reveal that the quarter-hour minute, used as a forecast confidence input in the proposed method, governs the relative contribution of forecasts (MPC actions) and real-time balancing data to the final decision. The proposed agent learned to primarily follow the MPC action at the beginning of the quarter hour, while increasingly relying on real-time balancing data to deviate from the MPC actions as the quarter hour progresses.


\printbibliography
\end{document}